\documentstyle[amssymb,prb,aps,psfig,floats]{revtex}

\twocolumn

\begin{document}

\draft

\wideabs{

\title{Hopping conductivity in heavily doped n-type GaAs layers in the quantum Hall
effect regime}
\author{S. S. Murzin$^{\text{1,2}}$, M. Weiss$^{\text{1}}$, A. G. M. Jansen$^{\text{1%
}}$ and K. Eberl$^{\text{3}}$}
\address{$^{\text{1}}$Grenoble High Magnetic Field Laboratory, Max-Plank-Institut\\
f\"{u}r Festk\"{o}rperforschung and Centre National de la Recherche\\
Scientifique, BP 166, F-38042, Grenoble Cedex 9, France \\
$^{\text{2}}$Institute of Solid State Physics RAS, 142432, Chernogolovka,\\
Moscow District., Russia\\
$^{\text{3}}$Max-Plank-Institut f\"{u}r Festk\"{o}rperforschung, Postfach\\
800 665 D-70569, Stuttgart, Germany}

\maketitle

\begin{abstract}
We investigate the magnetoresistance of epitaxially grown, heavily doped
n-type GaAs layers with thickness (40-50 nm) larger than the electronic mean
free path (23 nm). The temperature dependence of the dissipative resistance $%
R_{xx}$ in the quantum Hall effect regime can be well described by a hopping
law $(R_{xx}\propto \exp {\{-(T_{0}/T)}^{p}\})$ with $p\approx 0.6$. We
discuss this result in terms of variable range hopping in a Coulomb gap
together with a dependence of the electron localization length on the energy
in the gap. The value of the exponent $p\geqslant 0.5$ shows that
electron-electron interactions have to be taken into account in order to
explain the occurrence of the quantum Hall effect in these samples, which
have a three-dimensional single electron density of states.
\end{abstract}

}

\narrowtext
For a two-dimensional electron system it is well known that the discrete
electron spectrum in a high magnetic field leads to quantized Hall
resistance (Quantum Hall Effect). However, Landau quantization is not a
strict prerequisite for the QHE. According to gauge arguments \cite
{Laugh,AoAn} it is sufficient that the dissipative conductance $G_{xx}$
vanishes at the Fermi level, and that delocalized states exist below. The
occurrence of the quantum Hall effect in not strictly two-dimensional
systems has been considered by Khmelnitzkii \cite{Khm} in conjunction with
the scaling theoretical treatment of the QHE\cite{Pruisken}.

In our previous works \cite{MJL,MCJ}, we observed the quantum Hall effect in
a strongly disordered system, which consisted of a heavily Si-doped (n-type)
GaAs layer between undoped GaAs. In this system a wide, smooth quantum well
is formed by the impurity space charge potential that builds up at the layer
interfaces. The electron gas is therefore confined inside the heavily doped
GaAs layer, in the area of maximum disorder. The thickness $d$ of the layers
ranging from $50$ up to $140$ nm was larger than the electronic mean free
path $l$ of $15-30$ nm. The density of states (DOS) of noninteracting
electrons in these samples is therefore expected to be practically
three-dimensional. As the very strong disorder broadening in the samples
leads to a rather smooth density of states without the formation of gaps
between Landau levels even at the highest magnetic fields ($\approx 20$ T),
we have proposed a reduction of the diagonal conductance $G_{xx}$ due to
electron-electron-interaction effects in diffusive transport as a possible
explanation for the observed quantization of $R_{xy}$ in the investigated,
strongly disordered systems. Evidence for this explanation comes from the
temperature dependence in the quantum Hall minima of $G_{xx}$ in samples
with a thickness between $50$ and $140~$nm \cite{MCJ}, which is logarithmic
with temperature $T$, and thus resembles the temperature dependence that is
caused by quantum corrections due to electron-electron interactions in
disordered conductors both in weak \cite{AA} and in high magnetic fields
\cite{Hough,Gir}. However, the logarithmic decrease of $G_{xx}$ that was
found in Ref. \cite{MCJ} exceeds in amplitude the range, where the theory of
quantum correction \cite{AA,Hough,Gir} is applicable. Furthermore, in the
thinnest sample (with a layer thickness $d=50$ nm), which showed a fully
developed quantization of the Hall effect, the conductance deviated from the
logarithmic temperature dependence at the lowest temperatures for values of
B where $G_{xx}\rightarrow 0$ and where the Hall conductance $G_{xy}$
correspondingly shows a plateau at a value of $2e^{2}/h$.

In the current work we have investigated an additional number of strongly
disordered GaAs layers with smaller values of $d$ (namely $40$ and $50$ nm),
showing a fully developed QHE below $100$ mK. The obtained hopping law for
the temperature dependent dissipative resistance $R_{xx}$ is discussed in
terms of the opening of a Coulomb gap.

The Hall conductance quantization in the aforementioned,
quasi-three-dimensional systems with a ''bare'' (high temperature)
conductance $G_{xx}^{0}\gg e^{2}/h$ can be understood qualitatively in the
following way. Usually, in systems with coherent diffusive transport the
dissipative conductance $G_{xx}$ decreases with temperature $T$ due to
quantum corrections. The weak localization (single-particle) corrections are
suppressed in a magnetic field $B$ and reduce to \cite{Efetov}
\[
G_{xx}(L_{\varphi })=G_{xx}^{0}-\frac{2}{\pi ^{2}}\frac{e^{4}}{%
h^{2}G_{xx}^{0}}\ln (L_{\varphi }/L_{0})=
\]

\begin{equation}
G_{xx}^{0}-\frac{m}{\pi ^{2}}\frac{%
e^{4}}{h^{2}G_{xx}^{0}}\ln (T_{1}/T).  \label{ln1}
\end{equation}
Here $L_{\varphi }=\sqrt{D_{xx}^{0}\tau _{\varphi }}$ is the distance an
electron moves diffusively during the phase breaking time $\tau _{\varphi
}\propto T^{-m}$, $D_{xx}^{0}$ is the ''bare'' high-temperature diffusion
coefficient,
\begin{equation}
L_{0}=d\sqrt{D_{xx}^{0}/D_{zz}^{0}}=\sqrt{dG_{xx}^{0}/\sigma _{zz}}
\label{L_0}
\end{equation}
is the electron displacement in the plane of the layer (perpendicular to the
magnetic field) for the time of its diffusion across the layer (along the
magnetic field), $D_{zz}$ and $\sigma _{zz}$ are the diffusion coefficient
and the conductivity in the direction parallel to the magnetic field, $T_{1}$
is defined from the equation $d\approx \sqrt{2D_{zz}^{0}\tau _{\varphi
}(T_{1})}$. At low temperatures the phase breaks due to electron-electron
interactions, leading to $m=1$. The second order corrections in a magnetic
field (Eq.\ref{ln1}) are much smaller ($\pi hG_{xx}^{0}/e^{2}$ times) than
the first-order corrections in zero field. Nevertheless, $G_{xx}$ will
eventually vanish, and in this case the Hall conductance $G_{xy}$ should be
quantized \cite{Laugh,AoAn}. Since $G_{xy}$ tends to different quantum
values for different bare Hall conductances $G_{xy}^{0}$, transitional
values of the bare conductance $G_{xy}^{0}$ should exist, for which $G_{xx}$
tends to a finite value and $G_{xy}$ is not quantized.

This approach, initially developed for spinless noninteracting electrons,
can give a reasonable, qualitative explanation for the occurrence of the
quantum Hall effect with even numbers of quantization $i$ in the above
mentioned, strongly disordered GaAs layers \cite{MJL,MCJ}. Quantitative
agreement with theory however does not exist because the quantum corrections
(Eq.\ref{ln1}) are small at real experimental conditions. To explain our
results, we have proposed the inclusion of electron-electron interactions.
In this case, the single-particle DOS and the conductance should decrease
with decreasing temperature due to quantum corrections caused by
interactions
\begin{equation}
G_{xx}(L_{T})=G_{xx}^{0}-\frac{\lambda e^{2}}{\pi h}\ln (T_{2}/T)=G_{xx}^{0}-%
\frac{2\lambda e^{2}}{\pi h}\ln (L_{T}/L_{0})  \label{ln2}
\end{equation}
that occur both in weak \cite{AA} and in high magnetic fields \cite
{Hough,Gir}. Here $L_{T}\sim (D_{xx}^{0}\hbar /k_{B}T)^{1/2}\ $, $k_{B}\ $
is the Boltzmann constant, and $T_{2}\sim \hbar D_{zz}^{0}/k_{B}d^{2}$. $%
\lambda \leqslant 1$ is the constant of interaction, which is of the order
of unity and even somewhat larger in high magnetic fields
\mbox{$(\mu _{B}gB/k_{B}T\gg
1)$} than in zero field ($\mu _{B}$ is the Bohr magneton). For $%
G_{xx}^{0}\gg e^{2}/h$\ these corrections are much larger than the
single-particle localization contributions (Eq.\ref{ln1}). The interaction
corrections (Eq.\ref{ln2}) will lead to a vanishing of the dissipative
conductance $G_{xx}$ as a consequence of the opening of a Coulomb gap in the
single particle DOS. Since also in this scenario $G_{xx}$ will vanish at
zero temperature, the Hall conductance should be quantized.

The samples used were prepared by molecular-beam epitaxy: on a GaAs (100)
substrate the following layers were successively grown: an undoped GaAs
layer (0.1 $\mu $m), a periodic structure of 30 $\times $ GaAs/AlGaAs(10/10
nm), an undoped GaAs layer (0.5 $\mu $m), the heavily Si-doped GaAs layer
with a nominal thickness of $d=40$ (sample~40) and $50$ nm (sample~50) and
donor(Si) concentrations of $1.5\times 10^{17}$ cm$^{-3}$, and last a cap
layer of 0.5 $\mu $m GaAs (undoped). Samples with Hall bar geometries of a
width of 0.2~mm and a length of 1.4~mm were etched out of the wafers. A
phase sensitive ac-technique was used for the magnetotransport measurements
down to 80~mK. In the experiments the applied magnetic field of up to 15~T
was directed perpendicular to the layers. Samples from the same wafer showed
identical behavior. The electron densities per square as derived from the
slope of the Hall resistance $R_{xy}$ in weak magnetic fields ($0.5-3$ T) at
$T=4.2$ K are $N_{s}=4.5$ and $5.1\times 10^{11}$ cm$^{-2}$. The ''bare''
mobilities $\mu _{0}$ are equal to $2500$ and $2300$ cm$^{2}$/Vs for sample
40 and 50 respectively, and the electron mean free path is about $23$ nm for
both samples. For the calculation of $\mu _{0}$ we took the value of the
bare resistance $R_{0}$ in the point of intersection of the curves $%
R_{xx}(B) $ for different temperatures at $B=3.4$ T, taking into account
that the classical resistance does not depend on field.

\begin{figure}[t]
\psfig{figure=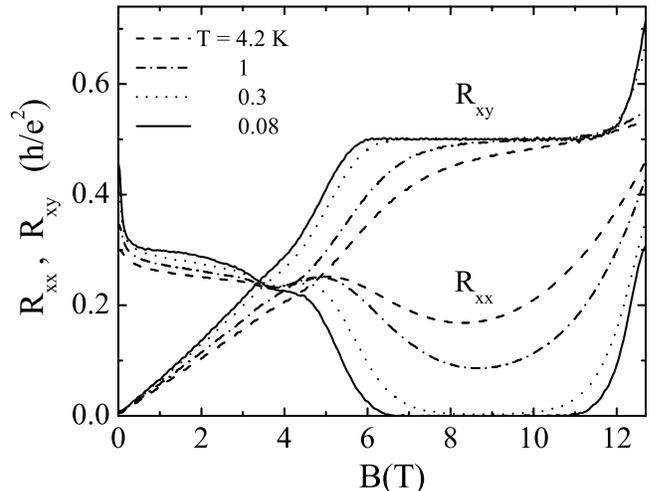,width=8.5cm}
\caption{Magnetic field dependence of the Hall ($R_{xy}$) and transverse ($%
R_{xx}$) resistance (per square) for sample~40 in a magnetic field
perpendicular to the heavily doped GaAs layer at different temperatures}
\label{R40}
\end{figure}

In Fig.\ref{R40} the magnetotransport data, namely the Hall ($R_{xy}$) and
transverse ($R_{xx}$, per square) resistance are plotted for sample 40 at
temperatures below 4.2 K. The diagonal resistance $R_{xx}$ decreases sharply
at low magnetic fields due to the suppression of the weak localization
corrections, and continues to decrease slightly between $0.5$ and $4$ T. It
shows a deep minimum ranging from $6$ to $11$ T. The Hall resistance $R_{xy}$
shows a linear increase up to $5$ T, and then reveals a wide plateau from $%
B=6$ T up to $11$ T at the lowest temperatures with the value $%
R_{xy}=h/2e^{2}$ (i.e. $i=2$), in the same field range where $R_{xx}$ shows
a deep minimum.

The Hall conductance $G_{xy}=R_{xy}/(R_{xx}^{2}+R_{xy}^{2})$ \ in the field
range of $B=0.5-4$ T does not depend on temperature. The diagonal
conductance (per square) $G_{xx}$ however shows a logarithmic temperature
dependence with an only slightly field dependent coefficient, while the
value of $G_{xx}$ itself changes considerably. This behavior is in agreement
with equation (\ref{ln2}), giving an interaction constant $\lambda \approx
0.5$. The magnetotransport data for sample~50 are similar to the data for
sample~40.

In our previous investigations of identical samples \cite{MJL,MCJ,Mcc} with
however a larger layer thickness, we found corrections to the conductivity
due to electron-electron interactions. In a region of low magnetic field ($%
B<4$ T) where $G_{xx}^{0}\gg e^{2}/h$ the magnetoresistance data can be
quantitatively described in terms of quantum corrections due to
electron-electron-interaction effects \cite{Mcc}. In high magnetic fields,
even in samples with thicknesses $d$ ranging up to $140$ nm, quantization of
the Hall conductance is observed. The mentioned samples show values of the
bare conductance $G_{xx}^{0}$ up to $2.6e^{2}/h$ \cite{MJL}. Even at these
high fields the different QHE minima in the transverse conductance $G_{xx}$
of different samples show a universal logarithmic temperature dependence in
a large range of a rescaled temperature $T/T_{sc}$, where $T_{sc}\varpropto
\exp (-3G_{xx}^{0}h/e^{2})$ \cite{MCJ}. Note however, that the decrease of $%
G_{xx}$ is not small and that a logarithmic temperature dependence is
observed beyond the region of applicability of the theory of quantum
corrections \cite{AA}. In the thinnest sample ($d=50$ nm) investigated in
Ref. \cite{MCJ}, showing a well pronounced QH plateau, a deviation from the
logarithmic behavior becomes visible at the lowest temperatures ($T<1$ K).
It is this range of temperature and layer thickness, that the present work
is focused on. We therefore study the temperature dependence of the
resistance $R_{xx}$ of samples with a thickness $d\leq 50$ nm, and therefore
a rather low bare conductance $G_{xx}^{0}$ of about $e^{2}/h$. These samples
show a pronounced plateau in $R_{xy}$ and a strong T-dependence near the
minimum of $R_{xx}$ at low temperatures, as shown in Fig. \ref{R40}.

\begin{figure}[t]
\psfig{figure=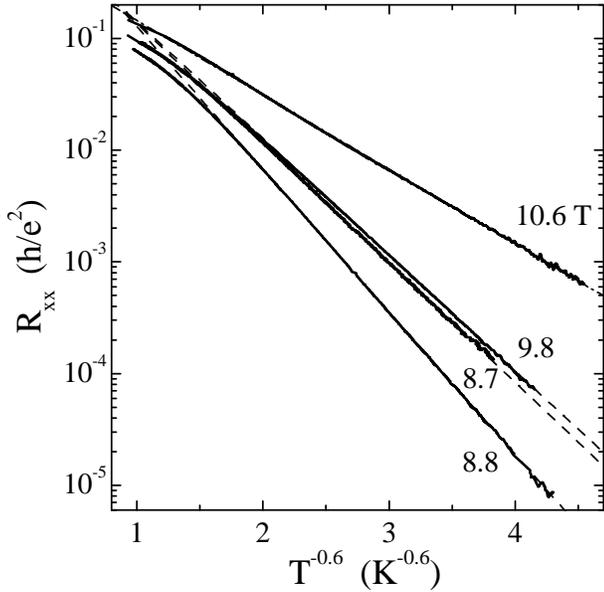,width=8cm}
\caption{The logarithm of the
resistance $R_{xx}$ as a function of $T^{-0.6}$ for sample 40 in
the minimum ($B=8.8$ T) and at larger fields indicated by lines,
and for sample 50 in the minimum ($B=8.7$ T)}
\label{06}
\end{figure}

In Fig. \ref{06} we plot the logarithm of the resistance $R_{xx}$ as a
function of $T^{-0.6}$ in the minima of $R_{xx}$ corresponding to the
plateaus at $R_{xy}=h/2e^{2}$ for samples 40 and 50 \cite{a}, and
additionally for sample 40 at somewhat larger $B$, but still not far from
the minimum. The exponent $p=0.6$ is chosen as a result of a fit of the
experimental data to a hopping law
\begin{equation}
R_{xx}=R_{0}\exp {\{-}(T_{0}/T{)}^{p}\}  \label{exp}
\end{equation}
in a range of temperature where $R_{xx}(T)<0.1R_{xx}(4.2$ K$)\approx
0.02h/e^{2}$. The fitting parameters $R_{0}$ and $T_{0}$ are listed in the
table.

Attempts to fit the data by an expression with a temperature dependent
prefactor
\begin{equation}
R_{xx}=\alpha T^{r}\exp {\{-}(T_{0}/T{)}^{p}{\}}  \label{prefactor}
\end{equation}
and a fixed $p$ different from 0.6 resulted in a less optimal fit. Moreover,
the resulting fitting parameters are unphysical. For instance, for the case
of $p=0.5$ the fit gives $r=0.65,$ $\alpha =17.1$ and $T_{0}=25.5~$K. For
this situation, the prefactor $\alpha T^{r}$ in equation (\ref{prefactor})
at $T=1$K corresponds to a conductance $%
G_{xx}=R_{xx}/R_{xy}^{2}=R_{xx}/(0.5h/e^{2})^{2}\approx 70e^{2}/h$ while $%
G_{xx}=0.95e^{2}/h$ only at \mbox{$T=10$ K.} The large prefactor in the
conductance is compensated by a small exponential factor $\exp {\{-(T}_{0}{%
/T)}^{p}\}=\exp {\{-25.5}^{1/2}\}\approx 6.4\times 10^{-3}$, while $%
G_{xx}(10 $K$)/G_{xx}(1~$K$)$ has a value of about $3$ only. The small
difference between $G_{xx}(1~$K$)$ and $G_{xx}(10$K$)$ would be the result
of a compensation of the two, which is not realistic. Thus we conclude, that
the temperature dependence in the $i=2$ minimum in $R_{xx}$ is rather
described by a hopping law according to Eq.(\ref{exp}) with a hopping
exponent $\ p$ near $0.6$.

Without the existence of a Coulomb gap the Mott theory of variable range
hopping \cite{Mott} predicts the temperature dependence of $R_{xx}$ to
follow equation (\ref{exp}) with $p=1/3$. According to the theory from Efros
and Sklovskii \cite{ESh,ESh2}, $p$ is equal to $1/2$ in the presence of a
Coulomb gap around the Fermi energy $E_{F}$ (both in zero magnetic field and
in the QHE regime). This theory was developed for situations where the
localization length $\xi $\ does not depend on the energy
\mbox{$\epsilon
=|E-E_{F}|$} in the gap. In the case of Anderson localization the
localization length $\xi $ should depend on the energy $\epsilon $ near the
Coulomb gap.

In the single-particle approach, at $G_{xy}^{0}(B)=ie^{2}/h$ with even $i$,
the localization length $\xi _{sp}$ of an electron at the Fermi level equals
\begin{equation}
\xi _{sp}\sim L_{0}\exp \left( 0.5\pi ^{2}G_{xx}^{0}~^{2}h^{2}/e^{4}\right)
\label{xi_inf}
\end{equation}
estimated from the equation $G_{xx}(\xi _{sp})=0$ with $G_{xx}$\ taken from
equation (\ref{ln1}). According to the scaling theoretical treatment of the
QHE, the localization length $\xi _{sp}$ generally depends both on $%
G_{xx}^{0}$ and $G_{xy}^{0}$. It diverges at $G_{xy}^{0}(B)=(i+1/2)e^{2}/h$.

However, electron-electron interactions should result in a decrease of the
localization length in the Coulomb gap. A lower limit of this decrease can
be estimated from the equation $G_{xx}(\xi _{0})=0$ with $G_{xx}$\ taken
from equation (\ref{ln2})
\begin{equation}
\xi _{0}\sim L_{0}\exp \left( \frac{\pi G_{xx}^{0}~h}{2\lambda e^{2}}\right)
\label{xi_0}.
\end{equation}
Outside the gap interaction is not important, and the localization length is
equal or larger than the one given by expression (\ref{xi_inf}) with $%
G_{xx}^{0}=G_{xx}^{0}(E)$ for the energy $E$. For typical values of $%
G_{xx}^{0}\approx e^{2}/h$ and $\lambda \approx 1$, $\xi _{0}$ is much
smaller than $\xi _{sp}$. As it will be shown below, such an energy
dependence of $\xi $ should result in $p>1/2$ in Eq.(\ref{exp}).

The single-particle density of states should be unaffected by an energy
dependence of the localization length, unless the distance between electrons
is much larger than the localization length, i.e. $g(\epsilon )\xi (\epsilon
)^{2}|\epsilon |\ll 1$. It should still be linear: \ $g(\epsilon )=\gamma
|\epsilon |$ with $\gamma =2\kappa ^{2}/\pi e^{4}$ ($\kappa $ is the
dielectric constant of the lattice). Let us suppose that in some range of
energy $\xi =\alpha |\epsilon |^{s}$. Then by analogy with the Mott-law
derivation \cite{Mott,ESh2} we obtain
\begin{equation}
R_{xx}\varpropto G_{xx}\propto \exp {\{-(T}_{0}{/T)}^{(s+1)/(s+2)}{\},}
\label{G}
\end{equation}
where
\[
T_{0}=\frac{S}{k_{B}}\left( \frac{Ce^{2}}{\alpha \kappa }\right)
^{1/(s+1)},
\]
\begin{equation}
S=\left[ (s+1)^{-(s+1)/(s+2)}+(s+1)^{1/(s+2)}\right]
^{(s+2)/(s+1)}.  \label{T0}
\end{equation}
The coefficient $\alpha $ depends on the magnetic field. For $s=0$ equations
(\ref{G}) and (\ref{T0}) reduce to the results from Efros and Shklovskii.
For $s\gg 1$ one finds activated behavior and $p=(s+1)/(s+2)=0.6$ is
obtained for $s=1/2$. The main contribution to the conductivity is given by
hopping electrons with an energy of
\[
\epsilon _{h}=\left[ \frac{Ck_{B}e^{2}(s+1)}{\alpha \kappa }T\right]
^{1/(s+2)}
\]
The corresponding localization length
\begin{equation}
\xi _{h}=\alpha |\epsilon _{h}|^{s}=\left[ \frac{Ck_{B}e^{2}(s+1)}{\alpha
\kappa }T\right] ^{s/(s+2)},
\end{equation}
of the electrons giving the main contribution to the conductivity for $T=0.1$
K and $s=1/2$ is listed in the table. The numerical coefficient $C$\ is
taken to be 1.55 as defined from the equation for $T_{0}$ from the
Efros-Shklovskii theory ($T_{0}=6.2e^{2}/\varepsilon \xi $) \cite{Nguen}.

\begin{table}[tbp]
\begin{tabular}{cccccc}
Sample & $B$ (T) & $T_{0}$ (K) & $R_{0}$ ($h/e^{2}$) & $\xi _{h}$ ($\mu m)$
& $\xi _{T}$ ($\mu m)$ \\
40 & 8.8 & 6.0 & 2.41 & 0.63 & 0.24 \\
40 & 9.8 & 4.3 & 1.55 & 0.95 & 0.4 \\
40 & 10.6 & 2.04 & 0.66 & 2.3 & 1.2 \\
50 & 8.7 & 4.5 & 1.6 & 0.9 & 0.37
\end{tabular}
\caption{Values of the magnetic fields $B$ , the constant $T_{0}$ and the
prefactor $R_{0}$, the localization length $\protect\xi _{h}$\ of the
electrons giving the main contribution to the conductivity, at $T=0.1$ K and
the localization length $\protect\xi _{T}$\ of the electrons with energy $%
\protect\epsilon /k_{B}=0.1$ K. }
\label{t}
\end{table}

Since $\xi _{h}\varpropto T^{1/5}$ can not be smaller than $\xi _{0}$ it
should become constant at the lowest temperatures and the temperature
dependence should reduce to the Efros-Shklovskii law. In our experimental
conditions $\xi _{h}$ approaches $\xi_{0}<200$ nm at the very small
temperature of $T<3\times 10^{-4}$ K at $B=8.8$ T and $T<5\times 10^{-7}$ K
at $B=10.6$ T for sample~40.

A dependence of the localization length on energy could probably also
account for hopping exponents $p>1/2$, observed in zero-field experiments
\cite{KHJD,Gershen,metals}. An energy-dependence as described above is also
indicated by numerical simulations \cite{Voj,JW}. Therefore, also in zero
magnetic field the power of $T$ in equation (\ref{exp}) could be larger
than $1/2$ in some range of temperature.

In summary, in low magnetic fields (but still larger than $0.5$ T) where $%
G_{xx}>3e^{2}/h$, the temperature dependence of the diagonal conductance $%
G_{xx}$ of heavily doped n-type GaAs layers with thicknesses ($d=40\div 140$ nm)
larger than the mean free path of the electrons ($l=23$ nm) is well
described by the theory of quantum corrections due to electron-electron
interactions. In high magnetic fields where $G_{xx}<3e^{2}/h$ the temperature
dependence of the conductance in the minima of $G_{xx,min}$ is still close to
logarithmic down to $0.25e^{2}/h$, although the theory of quantum corrections is
no more applicable. In the region of $G_{xx}<0.25e^{2}/h$ the dissipative
conductance shows an exponential decrease with a power $p\approx 0.6$,
indicating the presence of a Coulomb gap. The data display the relevance of
electron-electron interactions for the quantum Hall effect in these systems
which have a 3D single-particle spectrum.

We have pointed out, that a dependence of the localization length on energy
could result in an exponent $p>1/2$ both in zero and nonzero magnetic field.


\begin{references}
\bibitem{Efetov}  The numerical coefficient in Eq.(\ref{ln1}) differs from\
an earlier presented one \cite{Huck,Mcc}. Therefore, we have recalculated it
using the equation for the $\beta $-function derived by S. Hikami \cite
{Hikami} and K. B. Efetov \cite{Efetov2}, taking into account that the
equation is written for a spinless electron system.
\bibitem{a}  The very first measurement of sample~$50$ \cite{MCJ} showed
an activated temperature dependence below $0.3$ K. The results of subsequent
measurements, presented here and taken with special care to avoid heating
effects, are identical for two samples from the same wafer with $d=50$ nm.
\end{references}
\end{document}